\documentclass[sigconf,nonacm]{acmart}
\usepackage{booktabs}  
\usepackage{multirow}  

\usepackage{graphicx}  

\usepackage{balance}   
\usepackage{tikz}
\usepackage{pgfplots}
\pgfplotsset{compat=1.17}
\usetikzlibrary{shapes.geometric, arrows.meta, positioning, calc, fit, backgrounds, decorations.pathreplacing}

\AtBeginDocument{%
  }

\begin{document}

\title{ECHO: Explainable Co-editing with Human-in-the-loop Operations for Presentation Refinement}

\author{Yu Fu}
\email{fuyu05@stu.scu.edu.cn}

\affiliation{%
  \institution{Sichuan University}
  \city{Chengdu}
  \state{Sichuan}
  \country{China}
}

\author{Yongqi Kang}
\email{2023141520239@stu.scu.edu.cn}

\affiliation{%
  \institution{Sichuan University}
  \city{Chengdu}
  \state{Sichuan}
  \country{China}
}

\author{Yujia Zhou}
\email{2023141520192@stu.scu.edu.cn}

\affiliation{%
  \institution{Sichuan University}
  \city{Chengdu}
  \state{Sichuan}
  \country{China}
}

\author{Yong Zhao}
\email{yong.zhao@scupi.cn}

\authornotemark[1]
\affiliation{%
  \institution{Sichuan University}
  \city{Chengdu}
  \state{Sichuan}
  \country{China}
}

\renewcommand{\shortauthors}{Fu et al.}

\begin{abstract}
  Authoring and refining presentation slides is a highly time-consuming core task in academic and business domains. While generative AI tools have lowered the barrier for creating initial drafts, their "black-box, one-way generation" paradigm severely deprives users of fine-grained control. Through a formative study (N=10), we identified "trial-and-error anxiety" and "inconsistent cross-page formatting" as primary bottlenecks in human-AI co-creation. Consequently, we present ECHO, an interactive system based on multimodal intent grounding and explainable operation plans. ECHO enables precise local edits via a "natural language + visual selection" paradigm, utilizing a decoupled "Plan-Confirm-Execute" loop and dynamic memory mechanisms to transform implicit AI intents into highly controllable layout co-creation.
 
    To systematically evaluate document refinement, we propose the CoEdit-Eval framework. Objective evaluations across multiple foundation models (e.g., GPT-5, GLM-4.7) demonstrate that while baselines uniformly fail in intent mapping (0\% accuracy) and spatial grounding (0\% Hit@1), the ECHO architecture boosts Target Hit@1 to 55\%--85\% depending on the base model. Furthermore, integrating Vision-Language Models (VLMs) effectively resolves spatial ambiguities---achieving significant win rates in LLM blind evaluations---and our Undo mechanism guarantees 100\% physical file consistency (MD5 hash). Finally, a controlled study with 14 participants shows that ECHO significantly reduces cognitive workload (NASA-TLX scores dropped by 20.8\%, from 82.6 to 65.4) and reveals the dynamic evolution of human control allocation across different cognitive tasks.
\end{abstract}

\ccsdesc[500]{Human-centered computing~Interactive systems and tools}
\ccsdesc[500]{Human-centered computing~Natural language interfaces}
\ccsdesc[300]{Human-centered computing~Empirical studies in HCI}

\keywords{Human-AI Collaboration, Presentation Authoring; Generative AI, Human-in-the-Loop, Explainable Interfaces, Multimodal Interaction}


\maketitle

\begin{figure*}[t]
  \centering
  \includegraphics[width=\textwidth]{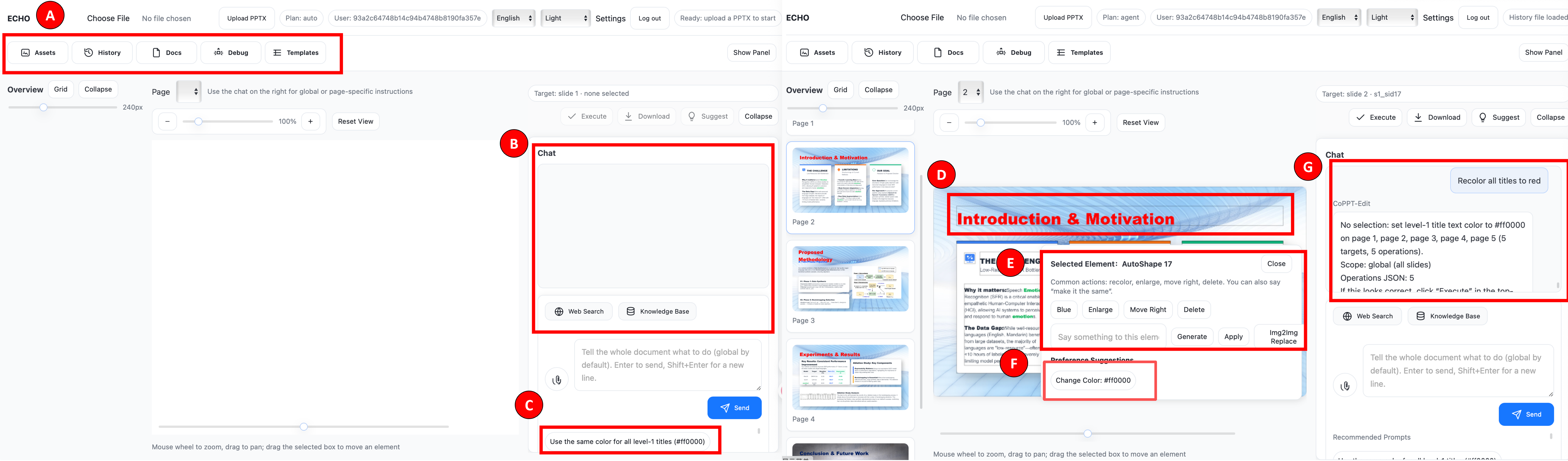}
  \caption{The ECHO workbench interface and a walkthrough of the multimodal editing workflow.
  \textbf{(A)}~\emph{Resource Library}: comprises five panels --- Assets (a repository of text-to-image and image-to-image outputs), History (previously uploaded documents), Docs (reference materials that can inform content edits), Debug (displaying the current Explanation, Suggestion, Operation JSON, and Selected Element), and Templates (user-uploaded style templates for one-click global restyling).
  \textbf{(B)}~\emph{Chat Interface}: the primary conversational channel through which users issue both global instructions (e.g., ``apply a dark theme to all slides'') and element-specific edits by referencing a selected target.
  \textbf{(C)}~\emph{Personalized Action Recommendations}: context-aware operation suggestions generated from the user's editing history and common formatting needs.
  \textbf{(D)}~\emph{Element Dialog}: clicking any element on the canvas opens an inline dialog box that accepts the same natural-language instructions as the chat interface, enabling rapid in-situ micro-edits.
  \textbf{(E)}~\emph{Habit Memory Recall}: because the user previously issued ``Recolor all titles to red'' (\texttt{\#ff0000}) via the element dialog, the system surfaces a persistent suggestion to apply the same color, demonstrating the User Habit Memory mechanism.
  \textbf{(F)}~\emph{Instruction Input}: the user types ``Recolor all titles to red'' in the element dialog.
  \textbf{(G)}~\emph{Execution Result}: the rendered slide after the instruction is confirmed and executed, showing all titles updated to \texttt{\#ff0000}.}
  \label{fig:echo-ui}
  \Description{Annotated screenshot of the ECHO interface showing seven labeled components (A--G): the resource library with five tabs, the main chat panel, personalized action recommendations, an element-level dialog box, a habit-memory suggestion, the instruction input field, and the resulting slide with red titles.}
\end{figure*}

\section{Introduction}
\label{sec:introduction}

Presentations serve as the core medium for visual communication, playing an indispensable role in scenarios ranging from academic reporting to commercial pitches. In recent years, the breakthroughs of Large Language Models (LLMs) have catalyzed numerous ``one-click'' presentation generation AI tools, significantly lowering the barrier for creating initial drafts \cite{zheng2025pptagent, ge2025autopresent}. However, this highly automated workflow often comes at the severe cost of depriving creators of their fine-grained control and agency.

To ground our design in real-world practice rather than in assumed needs, we conducted a formative study with 10 frequent presentation authors (5 graduate students preparing academic talks and 5 industry professionals producing commercial pitch decks). Section~\ref{sec:formative} reports the study in full; here we summarize the central finding that motivates ECHO: current AI-assisted tools predominantly adopt a ``black-box, unidirectional generation'' logic~\cite{zhang2025ufo}, in which the user issues a prompt and receives a finished slide, with no visibility into the intermediate reasoning and no channel for fine-grained intervention. Participants consistently described this regime as producing ``trial-and-error anxiety'' --- they could neither reliably steer the model toward a specific layout nor recover quickly when a generated draft drifted from their intent. When compounded with the need for visual consistency across a multi-slide deck, these frictions pushed participants to abandon AI assistance on mature documents and fall back on manual editing, precisely at the stage where the productivity gains would be most valuable.

To bridge this semantic gap and break the bottleneck of control, we designed and propose the \textbf{ECHO} system. ECHO is not merely an engineering artifact pursuing underlying generation accuracy; rather, it aims to explore the ideal boundaries of human-AI collaboration in complex document authoring \cite{vatavu2024ai}:

\begin{itemize}
    \item \textbf{Multimodal Intent Grounding and Explainable Operations:} The system constructs a composite interaction paradigm of ``Natural Language + Visual Selection (Target ID),'' allowing users to precisely specify local layers. More importantly, before executing any modification, ECHO translates the implicit semantic understanding into a structured operation plan (JSON Operations). These plans must undergo ``Human Validation'' before rendering, effectively shattering the black box of blind modifications \cite{liang2022multiviz}.
    
    \item \textbf{Dynamic Context and Personalized Memory Mechanism:} Addressing the strong demands for ``layout consistency'' and ``reusability'' expressed by interviewees, ECHO innovatively introduces long- and short-term memory mechanisms. The system can not only establish a \textit{``Global Style Memory''} through template extraction (solidifying the user's custom background, color, and font specifications) but also implicitly track and form a \textit{``User Habit Memory''} (recording high-frequency formatting preferences and historical edit actions in the current session). This bidirectional memory injection enables the AI to provide highly personalized editing suggestions tailored to the user's specific habits across multi-page, multi-turn interactions, truly achieving a leap from a ``stateless generator'' to an ``empathic layout collaborator.''
\end{itemize}

Figure~\ref{fig:echo-ui} provides an overview of the ECHO workbench and illustrates these mechanisms in action. The interface integrates a resource library, a conversational chat panel, and an element-level dialog into a unified canvas, allowing users to seamlessly combine global instructions with point-and-click micro-edits. Notably, the system surfaces personalized action recommendations derived from the user's editing history (e.g., recalling a previously applied color preference), making the Habit Memory mechanism tangible and immediately useful in practice.

The core contributions of this paper are fourfold:
\begin{itemize}
    \item \textbf{Empirical Insights:} We clarify the real-world pain points of human-AI collaboration in automated multimedia document generation through a formative study, distilling core design principles for controllable and personalized slide authoring tools.
    \item \textbf{System Architecture:} We propose and implement ECHO, a collaborative system integrating multimodal intent capture, an explainable human-in-the-loop execution loop, and dynamic habit/style memory, effectively bridging the gap in fine-grained and consistency control for complex document editing.
    \item \textbf{Novel Evaluation Methodology:} Addressing the limitations of existing benchmarks that heavily favor generation over editing, we propose \textbf{CoEdit-Eval}, a systematic evaluation framework tailored for slide editing. Combining objective engineering metrics with visual LLM-as-a-judge blind evaluations \cite{jung2025talk}, we provide a standardized benchmark for the iterative refinement of multimedia documents.
    \item \textbf{Strategy Evolution Analysis:} Through objective baseline evaluations and a controlled user study (N=14), we validate the system's robustness and deeply reveal the dynamic evolution of human-AI collaborative strategies and the allocation of control when facing diverse cognitive demands (e.g., rigorous academic illustration vs. divergent commercial ideation).
\end{itemize}

\textbf{Paper Overview:} The remainder of this paper is organized as follows. Section~\ref{sec:related_work} reviews related work on presentation generation, GUI agents, and multimodal human-AI collaboration, and articulates how ECHO differs from the closest prior system, Talk to Your Slides~\cite{jung2025talk}. Section~\ref{sec:formative} reports the formative study and the four design goals it produced. Section~\ref{sec:system} details the architecture of the ECHO system, including intent parsing, task-adaptive routing, and memory mechanisms. Section~\ref{sec:user_study} presents the controlled user study ($N$=14), reporting both quantitative outcomes and a qualitative analysis of how users' collaboration strategies evolve across tasks. Section~\ref{sec:evaluation} introduces the CoEdit-Eval framework and reports objective evaluation results across multiple foundation models. Section~\ref{sec:discussion} discusses design implications, limitations, and future directions. Section~\ref{sec:conclusion} concludes.

\section{Related Work}
\label{sec:related_work}
We situate ECHO against four bodies of prior work: automated document-to-presentation generation (\S\ref{2.1}), multimodal interaction for creative tools (\S\ref{2.2}), interpretability and controllability in generative AI (\S\ref{2.3}), and collaboration models in human--AI co-creation (\S\ref{2.4}). Within \S\ref{2.1} we also provide a focused, point-by-point comparison against the closest prior system, Talk to Your Slides~\cite{jung2025talk}, to make the contribution boundary of ECHO explicit.

\subsection{Automated Document-to-Presentation Generation}
\label{2.1}
Automatically generating presentation slides from source documents has been a longstanding task at the intersection of natural-language processing and multimodal generation, aiming to transform long structured documents into slides with clear logical structure and reasonable layout. Early work adopts an end-to-end generation paradigm: DOC2PPT~\cite{fu2022doc2ppt} first formulates document-to-slide as a complete generation task and employs a hierarchical sequence model that jointly handles text summarization, sentence rewriting, and layout generation; DocPres~\cite{bandyopadhyay2024enhancing} proposes a multi-stage LLM+VLM architecture for long documents, improving cross-slide consistency through hierarchical summarization, outline generation, and multimodal retrieval; and Personaware-D2S~\cite{mondal2024presentations} introduces audience and duration constraints, using human preference learning to produce personalized decks (e.g., expert vs.\ novice, long vs.\ short).

A parallel line of work explores human--AI collaboration and accessibility in slide authoring. PaperBridge~\cite{zhang2025paperbridge} supports research narrative construction by helping authors explore alternative narrative structures through co-exploration with an LLM. Diffscriber~\cite{peng2022diffscriber} automatically detects changes between slide versions and generates natural-language descriptions of those changes, enabling visually impaired collaborators to participate in design review. Talk to Your Slides~\cite{jung2025talk} uses an LLM-based agent to support natural-language-driven editing of existing slides, though it does not generate presentations from source documents.

\paragraph{Positioning ECHO against Talk to Your Slides~\cite{jung2025talk}.} Among prior systems, Talk to Your Slides is the most directly comparable to ECHO, as both target the \emph{refinement} of existing slides rather than zero-shot generation from scratch. Both share the premise that the critical unmet need in human--AI slide authoring is iterative editing, not initial drafting. However, ECHO differs from Talk to Your Slides along four dimensions that we argue are essential for moving from a research demonstration to a practically controllable co-authoring system.

\textit{(1) Interaction paradigm --- verbal-only vs.\ multimodal grounding.} Talk to Your Slides treats editing as a pure text-to-action mapping: the user issues a natural-language command and the system parses it into a slide modification. This inherits the spatial-reference bottleneck we identify in our formative study (\S\ref{sec:formative}, Theme~4): users must verbally describe target elements (``the second bullet in the left column''), which is both laborious and error-prone in dense layouts. ECHO introduces an explicit \emph{``pointing-as-speaking''} modality in which a canvas click supplies an unambiguous \texttt{Target ID} that is fused with the textual instruction. This is not a cosmetic UI refinement but a fundamental disambiguation mechanism: spatial-reference tasks that are ambiguous in pure text are resolved deterministically when a \texttt{Target ID} is provided, and we fall back to a VLM only when no selection is available (\S\ref{sec:evaluation}, Experiment~2).

\textit{(2) Execution model --- direct action vs.\ Plan-Confirm-Execute.} Talk to Your Slides executes edits immediately upon parsing, placing the user in the familiar but brittle posture of post-hoc inspection and rollback. In the language of our formative study (Theme~1), this does not relieve trial-and-error anxiety; it merely displaces it from prompt-crafting to result-inspecting. ECHO instead externalizes the LLM's planned actions as a structured JSON operation plan that the user inspects \emph{before} any change to the document. This shifts the locus of control from reactive correction to proactive authorization, and it is this decoupling that our NASA-TLX results (a 20.8\% overall workload reduction and a 38.5\% reduction in the Frustration subscale; \S\ref{sec:user_study}) most directly reflect.

\textit{(3) Correctness guarantees --- free-form generation vs.\ schema-constrained operations.} Talk to Your Slides relies on the LLM to produce syntactically valid edit commands, with no architectural guarantee that malformed or semantically impossible commands will be rejected before execution. Our objective evaluation (\S\ref{sec:evaluation}, Experiment~1) shows that without schema constraints, even strong foundation models produce unrenderable outputs in a substantial fraction of cases. ECHO's \texttt{ops.schema.json} constraint acts as an Abstract Syntax Tree that forces the LLM's output into a finite, deterministically executable atomic operation set; we argue, and our data support, that it is this constraint --- not the base model's raw capability --- that is principally responsible for the grounding gains we report.

\textit{(4) Safety model --- soft undo vs.\ byte-exact rollback.} Talk to Your Slides and similar systems rely on application-level undo, which offers no guarantee against partial or corrupted document states following a failed or hallucinated edit. This is precisely the failure mode participants described in our formative study (Theme~5), and it is the reason users self-limit AI assistance on mature documents. ECHO commits an incremental file-level snapshot before every write and verifies rollback with an MD5 hash check, achieving 100\% undo consistency across stress tests with templates of up to 445 atomic operations (\S\ref{sec:evaluation}, Experiment~3). This transforms reversibility from a best-effort UI convenience into a verified physical property of the system.

\paragraph{Relation to generation-oriented systems and GUI agents.} A second strand of recent work focuses on fully automated slide \emph{generation} rather than editing: PPTAgent~\cite{zheng2025pptagent} generates presentations end-to-end from documents and proposes evaluation beyond text-to-slide metrics; AutoPresent~\cite{ge2025autopresent} generates structured visuals from scratch via LLM agents; and DOC2PPT~\cite{fu2022doc2ppt} and DocPres~\cite{bandyopadhyay2024enhancing} similarly target one-shot generation. While these systems represent substantial progress on the generation side, they operate in a fundamentally different collaboration regime: the user provides input once and receives a finished artifact, with limited or no mechanism for intervening in the generation process. Our formative study suggests this regime is a poor fit for professional authoring contexts, where the value of AI assistance lies specifically in iterative refinement of high-stakes documents. We therefore view ECHO as \emph{complementary to, not competitive with}, these generation systems: a practitioner could use AutoPresent or PPTAgent to produce an initial draft and then invoke ECHO for the iterative polishing phase where current tools fail.

A third line of work --- UFO~\cite{zhang2025ufo} and related GUI agents --- pursues end-to-end visual automation, treating the target application as a black box manipulated through screenshots. This approach is attractive for its generality but incurs substantial latency and cost, and it forgoes the strong correctness guarantees available when the document's internal structure can be parsed directly. Our task-adaptive routing result (\S\ref{sec:evaluation}, Experiment~2) provides empirical evidence for an intermediate position: pure-text reasoning over a parsed structure tree handles a large share of editing tasks with high accuracy and low cost, and VLM invocation is best reserved for the minority of cases with genuine spatial ambiguity. We view this as a broader architectural lesson for document-centric agents: \emph{full visual perception is not a precondition for controllable document editing, provided the document's native structure is exploitable}.

Table~\ref{tab:system-comparison} summarizes these distinctions along the dimensions surfaced by our formative study.

\begin{table*}[t]
\centering
\small
\caption{Comparison of ECHO against representative slide-authoring and document-agent systems along the dimensions surfaced by our formative study (\S\ref{sec:formative}).}
\label{tab:system-comparison}
\begin{tabular}{lllllll}
\toprule
\textbf{System} & \textbf{Task scope} & \textbf{Interaction} & \textbf{Pre-exec.\ plan} & \textbf{Correctness guarantee} & \textbf{Style/habit memory} & \textbf{File-level undo} \\
\midrule
DOC2PPT~\cite{fu2022doc2ppt}        & Generation      & None (batch)           & ---                   & None              & No                 & N/A \\
DocPres~\cite{bandyopadhyay2024enhancing} & Generation  & None (batch)           & ---                   & None              & No                 & N/A \\
AutoPresent~\cite{ge2025autopresent} & Generation     & None (batch)           & ---                   & Loose             & No                 & N/A \\
PPTAgent~\cite{zheng2025pptagent}    & Gen.\ + eval   & None (batch)           & ---                   & Loose             & No                 & N/A \\
Talk to Your Slides~\cite{jung2025talk} & Editing     & Text only              & No                    & Best-effort       & No                 & App-level \\
UFO (GUI agent)~\cite{zhang2025ufo}  & Generic visual & Screenshot + text      & Limited               & Best-effort       & No                 & App-level \\
\textbf{ECHO (ours)}                 & \textbf{Editing} & \textbf{Text + canvas click} & \textbf{Yes (JSON)} & \textbf{Schema-enforced} & \textbf{Global + habit} & \textbf{MD5-verified} \\
\bottomrule
\end{tabular}
\end{table*}

\subsection{Multimodal Interaction for Creative Tools}
\label{2.2}
Multimodal interaction allows users to steer a creative process more naturally by integrating language, vision, gesture, and touch. At the level of intent understanding, Zhang et al.~\cite{zhang2025multimodal} combine textual, visual, and audio signals to improve both classification accuracy and out-of-distribution detection of user intent; the Plug-and-Play Clarifier~\cite{yang2026plug} proposes a zero-shot multimodal framework for disambiguating egocentric user intent when language is underspecified or visual context is missing; and the survey by Dumas et al.~\cite{dumas2009multimodal} provides a foundational account of multimodal interface models and architectures.

At the level of creative tools, Narrative Motion Blocks~\cite{bourgault2025narrative} combines natural-language commands with direct drag-and-drop manipulation for fine-grained animation control; PlayWrite~\cite{tutuncu2026playwrite} integrates gesture and voice in XR to support collaborative narrative authoring; and Singh et al.~\cite{singh2023hide} use multimodal text--image--audio feedback to scaffold creative writing.

Most of this work, however, targets general-purpose creation or specific domains (animation, XR, writing), and rarely designs a lightweight, interpretable, and instantly-revisable multimodal interaction pattern for the specific task of slide refinement. Existing tools do not let users adjust a single element on a generated slide through a brief spoken or pointing instruction --- which is precisely the interaction scenario ECHO targets.

\subsection{Interpretability of Generative AI and Human-in-the-Loop Control}
\label{2.3}
Interpretability and controllability are central to ensuring that generated content is reliable, trustworthy, and usable. Moreno and Mart\'inez~\cite{moreno2026human} propose a two-layer human-in/on-the-loop framework for accessible text generation that improves factual correctness through staged human supervision. Ding~\cite{ding2024advancing} surveys the design space of human--AI interaction for generative GUIs, characterizing typical patterns such as guidance, selection, post-editing, and interactive revision.

On the theoretical side, the XAI survey by Ali et al.~\cite{ali2023explainable} reviews both intrinsically interpretable models and post-hoc explanation methods. In the human--AI co-creation literature, Oh et al.~\cite{oh2018lead} show empirically that user-led interaction, measured AI assistance, and process transparency all significantly improve trust; Scideator~\cite{radensky2024scideator} mitigates hallucination in scientific idea generation through structured guidance and human verification; and Zhou et al.~\cite{zhou2024understanding} show that nonlinear, iterative, and traceable collaboration patterns are better suited to reliable generation than one-shot pipelines.

Although controllable and interpretable generation has received wide attention, these techniques have rarely been integrated deeply into slide authoring. Users typically cannot inspect \emph{why} a particular slide was generated in a particular way, nor can they intervene in intermediate steps. ECHO targets exactly this gap by making the generation-to-edit transformation an interpretable, human-in-the-loop process: users not only see ``why the AI proposes this change,'' they authorize, reject, or amend it before it is applied.

\subsection{Collaboration Models and User Strategies in Human--AI Co-creation}
\label{2.4}
Understanding user behavior and creative strategy is central to designing co-creation systems that feel natural to use. Zhang et al.~\cite{zhang2025exploring} offer a scoping review of collaboration patterns, control distributions, and decision-making preferences reported in top-tier HCI venues. T2VTree~\cite{zheng2026t2vtree} supports branched, iterative creation through a tree-structured exploration interface. Moruzzi and Margarido~\cite{moruzzi2024user} propose a user-centered framework for human--AI co-creativity emphasizing user-led control, flexible division of labor, and adaptive interaction rhythm. Dodeja et al.~\cite{dodeja2024towards} take this a step further with strategy-level recommendations, enabling a system to dynamically adapt to user habits over time.

While these studies inform the design of user-centered co-creation systems broadly, most target open-domain writing, data exploration, or creative design; they rarely examine human--AI collaboration strategies for the structured, multi-stage task of slide authoring. Slide authoring in practice has distinct phase-dependent control demands --- coarse control over outline structure, tighter control over per-slide content allocation, and fine-grained control over typography and layout. ECHO is designed with this phased structure in mind and, through its explicit Plan-Confirm-Execute loop and traceable operation history, supports the dynamic reallocation of control across these phases. Our user study (\S\ref{sec:user_study}) empirically validates this design, showing that the same user flexibly shifts between tight structural control in academic tasks and looser exploratory control in creative tasks.

\section{Formative Study: Understanding Pain Points in Human--AI Slide Co-Creation}
\label{sec:formative}

To ground the design of ECHO in real-world practice rather than in speculative assumptions about user needs, we conducted a formative study aimed at characterizing how experienced presentation authors currently collaborate with generative AI, where the collaboration breaks down, and which design levers might most effectively unblock them. This section details the study methodology, participant population, the five themes distilled from our data, and the four design goals (DG1--DG4) that directly shaped the ECHO architecture.

\subsection{Method}
\label{subsec:formative-method}

\paragraph{Participants.} We recruited 10 participants (P1--P10; 6 female, 4 male; aged 22--38, $M$=27.4) through university mailing lists and professional referrals, using purposive sampling to balance two populations with distinct authoring demands: (i) 5 students (P1--P5) from computer science, design, and the life sciences who authored academic presentations (thesis proposals, conference talks, lab meeting presentations) at least twice per month; and (ii) 5 industry professionals (P6--P10) from consulting, product management, UX design, marketing, and venture capital who routinely produced commercial pitch decks, client reports, or internal strategy decks. All participants had used at least two AI-assisted authoring tools within the past six months (ChatGPT, Gamma, Tome, Microsoft Copilot, Canva Magic Design, Gemini, Doubao, Claude, kimi, AiPPT and Deepseek), ensuring that our findings reflect the experience of moderately sophisticated users rather than first-time adopters. Table~\ref{tab:formative-participants} summarizes participant profiles.

\paragraph{Procedure.} Each session lasted 35--55 minutes and was conducted remotely over video conferencing. The protocol had three parts. In a \emph{contextual inquiry} (approximately 10 minutes), participants shared their screen and walked through a recent real presentation they had authored with AI assistance, narrating their workflow, the prompts they issued, and the edits they ultimately performed manually. In a \emph{semi-structured interview} (approximately 25 minutes), we probed moments of breakdown, trust calibration, error-recovery strategies, and unmet needs. Finally, in a \emph{speed-dating design probe} (approximately 15 minutes), we presented low-fidelity sketches of four interaction alternatives --- one-shot generation, chat-only editing, direct manipulation combined with chat, and a plan-preview paradigm --- and elicited reactions. Sessions were audio-recorded and transcribed verbatim. The study was approved by our Institutional Review Board.

\paragraph{Analysis.} Two researchers independently performed reflexive thematic analysis~\cite{Braun08082019} on the transcripts. After open-coding the first four transcripts, the researchers met to consolidate an initial codebook (47 codes), which was iteratively refined across the remaining transcripts. Codes were clustered into candidate themes through affinity diagramming, reviewed against the raw data for internal coherence, and finalized through discussion. Inter-coder agreement on the final codebook reached Cohen's $\kappa$=0.81. Five themes and four derived design goals emerged.

\begin{table*}[t]
\centering
\small
\caption{Formative study participant profiles ($N$=10).}
\label{tab:formative-participants}
\begin{tabular}{lllll}
\toprule
\textbf{ID} & \textbf{Role} & \textbf{Domain} & \textbf{AI tools used in past 6 months} & \textbf{Authoring frequency} \\
\midrule
P1  & Undergraduate student          & Computer Science & ChatGPT, Gamma, AiPPT, kimi                  & Weekly \\
P2  & Master's student     & Computer Vision             & ChatGPT, Microsoft Copilot      & 2/month \\
P3  & PhD student          & Computational Biology       & ChatGPT, Tome, Gemini, Doubao                   & weekly \\
P4  & Undergraduate student          & Design Research             & ChatGPT, Canva Magic Design, SlidesAI     & Weekly \\
P5  & Undergraduate student     & Natural Language Processing & ChatGPT, Gamma, Deepseek, Claude, Doubao                  & 2/month \\
P6  & Product manager      & B2B SaaS                    & Qwen, Doubao, Deepseek, kimi         & Weekly \\
P7  & UX designer          & Fintech                     & Qwen, Canva Magic Design     & Weekly \\
P8  & Management consultant & Strategy consulting        & ChatGPT, Copilot                & Daily \\
P9  & Marketing lead       & Consumer goods              & Doubao, Gamma, Tome, AiPPT            & Weekly \\
P10 & Associate            & Early-stage venture capital & ChatGPT, Copilot, Claude, Gemini                & 2/month \\
\bottomrule
\end{tabular}
\end{table*}

\subsection{Findings}
\label{subsec:formative-findings}

\paragraph{Theme 1: Trial-and-error anxiety under opaque generation.}
All 10 participants described a recurring emotional arc when collaborating with current AI tools: an initial burst of optimism upon receiving a generated draft, followed by mounting frustration as successive prompt refinements failed to produce the intended layout. P3 described abandoning an AI-generated draft after ``the sixth prompt'': \emph{``I kept trying to say `move the chart up and make the title smaller,' and every time it would redo the whole slide, including parts I liked. Eventually I just gave up and dragged it by hand.''} The root cause participants articulated was not a lack of model capability but a lack of \emph{interaction granularity}: prompts operate at the whole-slide level, so even a small formatting tweak risks regenerating content the user has already approved. Seven participants (P1, P2, P3, P6, P7, P8, P10) explicitly used the word ``anxious'' or ``stressed'' to describe this state, and four reported that the anxiety escalated near deadlines, leading them to fall back on manual editing even when they believed the AI could, in principle, help.

\paragraph{Theme 2: The invisible-reasoning problem.}
Participants consistently struggled to form an accurate mental model of \emph{why} the AI made specific layout choices. P8, a management consultant, noted: \emph{``It's like asking an intern to redo a slide and they come back with something totally different --- but you can't ask them `what were you thinking?' You just get the output.''} This opacity produced two downstream problems. First, participants could not efficiently steer the AI because they did not know which part of their prompt was driving which part of the output. Second, trust was brittle: a single unexpected change --- even a minor one, like a shifted bullet indent --- caused participants to lose confidence in the entire slide and re-inspect every element. P4 captured this vividly: \emph{``One wrong font and I'm checking everything. It's more work than doing it myself.''} Notably, participants did not ask for full model interpretability; they asked for \emph{previewable, intervenable intermediate representations} of the AI's planned actions.

\paragraph{Theme 3: Cross-slide inconsistency and the burden of manual propagation.}
Eight of ten participants (all except P5 and P9, who tended to author shorter decks) reported that maintaining visual consistency across a multi-slide presentation was the single most time-consuming formatting task. Current tools handle individual slides competently but treat each generation as an isolated event, so font choices, color accents, heading alignment, and spacing subtly drift across slides. P6, a product manager, estimated spending roughly 40\% of her total authoring time on consistency repairs: \emph{``I'll have ten slides that each look fine on their own, and together they look like they were made by ten different people.''} Participants attempted workarounds --- copying a ``good'' slide as a template, manually propagating fonts --- but described these as tedious and error-prone. Notably, participants distinguished between two kinds of consistency needs: \emph{global style} (fonts, colors, margins --- amenable to template-level enforcement) and \emph{local pattern reuse} (e.g., ``always put the key takeaway in a colored pill in the top-right corner''), which current tools cannot capture at all.

\paragraph{Theme 4: Spatial reference is a language barrier.}
A subtle but pervasive pain point was the difficulty of referring to specific visual elements in natural language. Participants frequently produced long, awkward descriptions --- \emph{``the blue box on the left, below the pie chart, with the percentage in it''} (P2) --- only to have the AI edit the wrong element. P7 described resorting to screenshots and arrows: \emph{``I literally draw a red circle on a screenshot and paste it into ChatGPT. That's my life now.''} This friction was especially acute in dense layouts (dashboards, comparison tables) and when multiple elements shared similar attributes (e.g., three charts on one slide). Six participants (P1, P2, P4, P7, P9, P10) independently wished aloud for a \emph{``point-and-tell'' modality} where they could click an element and then speak about it, rather than verbally describing it.

\paragraph{Theme 5: Fear of irreversible damage.}
Four participants (P3, P5, P8, P10) described a specific anxiety around generative edits to documents containing significant prior work: \emph{``I don't trust it with a finished deck. What if it breaks a chart I spent two hours tuning?''} (P8). This fear was not abstract; three participants recounted specific incidents in which an AI edit had corrupted embedded content (a SmartArt diagram, an animation trigger, and a linked chart, respectively), and recovery had required re-authoring from scratch. As a consequence, participants self-limited their AI usage to drafting and low-stakes edits, explicitly avoiding AI assistance on mature documents --- exactly the stage where productivity gains would be most valuable. Participants uniformly expressed that a \emph{reliable, atomic undo} --- not merely a Ctrl+Z that might leave the document in a partially rewritten state --- would materially change their willingness to invite AI into late-stage editing.

\subsection{Derived Design Goals}
\label{subsec:formative-dgs}

Synthesizing across the five themes, we distilled four design goals that directly shaped ECHO's architecture. Each design goal maps to specific components of the system described in Section~\ref{sec:system}.

\begin{description}
    \item[DG1. Transparent intent.] Externalize the AI's planned actions as a reviewable, intervenable artifact \emph{before} execution. This addresses Themes 1 and 2 by converting implicit black-box reasoning into an explicit plan the user can accept, reject, or edit. Operationally, DG1 motivates ECHO's Plan-Confirm-Execute loop and its structured JSON-Operations representation (\S\ref{subsec:backend_engine}).

    \item[DG2. Multimodal grounding.] Replace verbal spatial description with direct visual selection. This addresses Theme 4 by letting users click elements on the canvas to supply an unambiguous target identifier, which is then fused with a natural-language instruction. DG2 motivates the \texttt{Target ID} mechanism and the task-adaptive VLM routing that activates only when no explicit selection is provided.

    \item[DG3. Persistent style and habit memory.] Treat the user's preferences as state that carries across turns and slides. This addresses Theme 3 by capturing both globally-applied style templates and implicitly-observed edit patterns, so the AI does not force the user to re-specify the same preferences on every slide. DG3 motivates ECHO's dual long-term (style) and short-term (habit) memory mechanisms.

    \item[DG4. Safe reversibility.] Guarantee physically exact rollback at the file level, not merely at the UI level. This addresses Theme 5 by committing an incremental document snapshot before every write, enabling byte-exact undo even after complex batch operations. DG4 motivates ECHO's deterministic execution engine with MD5-verified snapshotting.
\end{description}

These four design goals form the design rationale that the remaining sections validate in turn: Section~\ref{sec:system} describes the architecture, Section~\ref{sec:user_study} evaluates the user experience under controlled conditions, and Section~\ref{sec:evaluation} quantifies the objective performance of the system's core mechanisms.

\section{The ECHO System: Architecture and Multimodal Agent}
\label{sec:system}

To overcome the inherent drawbacks of traditional generative tools—such as ``black-box blind generation'' and the ``lack of local fine-grained control''—ECHO constructs a deeply integrated mixed-initiative authoring environment. Instead of treating Large Language Models (LLMs) as end-to-end black-box generators, the system reframes them as an intelligent agent equipped with perception, routing, and planning capabilities. Through an explicit \texttt{Plan-Confirm-Execute} workflow, ECHO enables highly controllable co-creation with a human-in-the-loop.

\begin{figure*}[t]
  \centering
  \includegraphics[width=\textwidth]{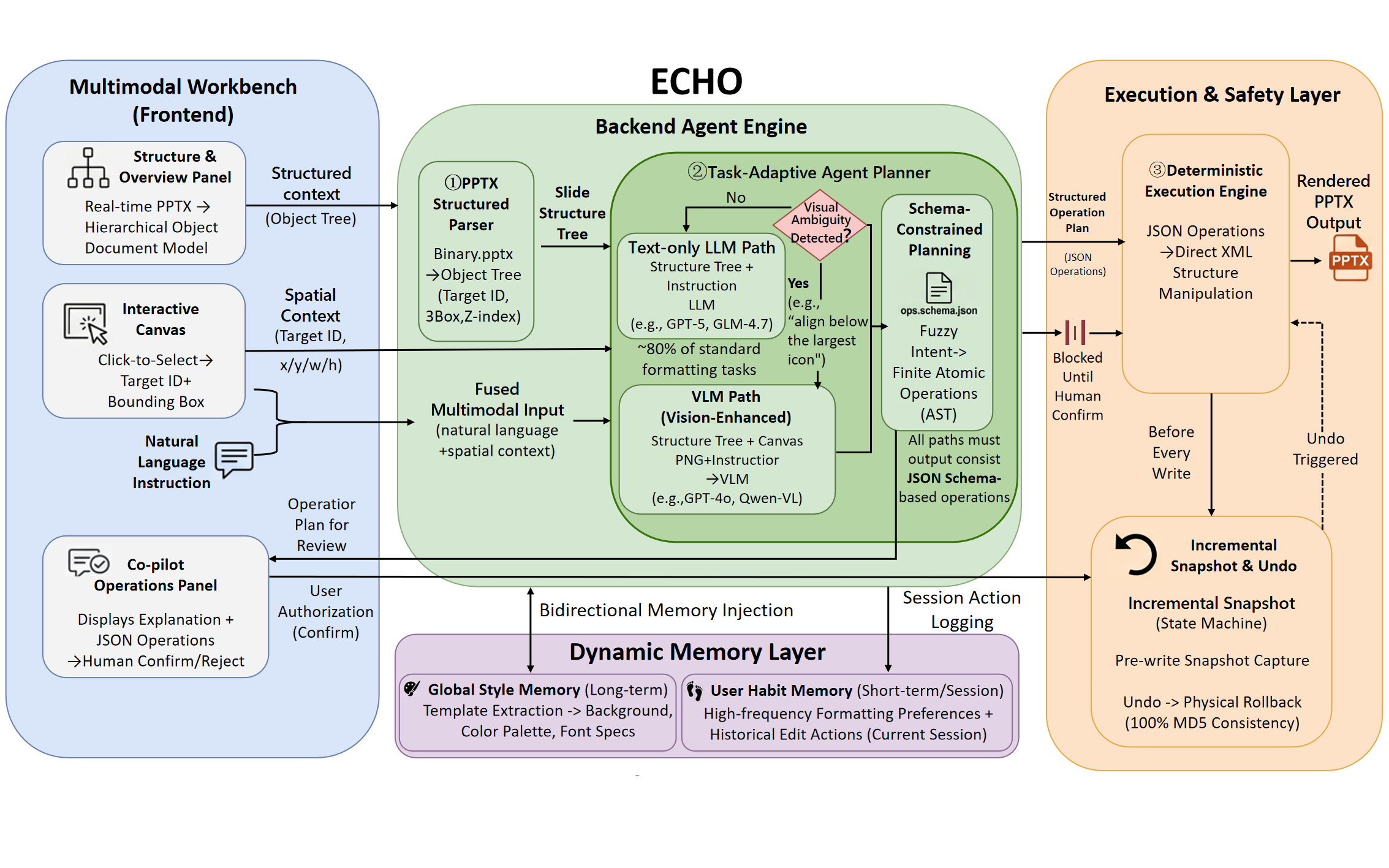}
  \caption{\textbf{The overall system architecture of ECHO.} The bidirectional workflow consists of three core modules: (1) \textbf{Multimodal Input \& Parsing}, which extracts the Document Object Tree and spatial Bounding Boxes to form structural context; (2) \textbf{Task-Adaptive Agent Planner}, which dynamically routes ambiguous spatial tasks to a Vision-Language Model (VLM) while utilizing a text-only LLM for strict Schema-Constrained Planning; and (3) \textbf{Deterministic Execution Engine}, which translates JSON operations into underlying XML manipulations, capturing incremental snapshots to guarantee physical undo safety before the final slide is rendered.}
  \label{fig:overall}
\end{figure*}

The Fig. \ref{fig:overall} shows overall structure of ECHO.

\subsection{Interface Overview and Multimodal Workbench}
The front-end of ECHO is not merely a presentation layer, but a multimodal context collector designed to seamlessly bridge human visual intuition with the Agent's semantic understanding. The workbench consists of three deeply interconnected modules:

\begin{itemize}
    \item \textbf{Structure \& Overview:} Serves as the macro-navigation anchor. The system parses complex PPTX binary files in real-time into a hierarchical Object Document Model (including text, images, and shapes). This provides the backend Agent planner with lightweight structured context and visual anchors.
    \item \textbf{Interactive Canvas:} Serves as the interface for visual feedback and multimodal spatial input. Users can directly point and click on specific element layers on the canvas. This action implicitly passes a unique, stable identifier (\texttt{Target ID}) and bounding box coordinates to the backend. This ``pointing-as-speaking'' modality endows the backend planner with precise ``Spatial Context,'' effectively resolving spatial referencing ambiguities.
    \item \textbf{Co-pilot Operations Panel:} Serves as a transparent window for the Agent's intents. Upon receiving user instructions, the system does not immediately manipulate the canvas. Instead, it presents a structured editing plan (comprising a natural language \texttt{Explanation} and specific JSON \texttt{Operations}) within the panel. The ``Confirm Execution'' button at the bottom establishes a strict boundary for control allocation, ensuring that all AI modifications must be explicitly authorized by the human.
\end{itemize}

\subsection{Usage Scenario Walkthrough: Human-AI Collaboration Strategy}
To illustrate how ECHO transforms tedious formatting tasks into fluid human-AI dialogues, we present a usage scenario involving Alice, a graduate student preparing a thesis proposal:
\begin{itemize}
    \item \textbf{Multimodal Intent and Precise Fine-Tuning:} Alice notices that a title is severely occluded by an experimental image. Instead of crafting a complex text prompt like ``move the blue text in the top-left corner,'' she directly clicks the title on the canvas and types: ``Move this above the image and change it to the accent color.'' The system instantly fuses the selected \texttt{Target ID} with the semantic instruction, generating an executable plan containing \texttt{move\_position} and \texttt{change\_color}. Once Alice clicks ``Confirm,'' the canvas renders a high-fidelity update instantly.
    \item \textbf{Long-Term Memory and Global Style Transfer:} When Alice wants the entire presentation to have a unified academic aesthetic, she enables the template function. The system retrieves the ``Global Style Memory'' and traverses all pages to generate batch update operations. This process strictly preserves the original images, charts, and layout logic, achieving a safe style transfer that changes the ``skin'' without damaging the ``bones.''
\end{itemize}

\subsection{Backend Engine and Bidirectional Analysis Architecture}
\label{subsec:backend_engine}
ECHO's fluid user experience and robust agent decision-making capabilities rely on a bidirectional analysis architecture that deeply integrates LLM reasoning with traditional document engineering. The backend pipeline is driven by three core engine modules:

\textbf{1. PPTX Structured Parser Engine:} 
Serving as the foundation of document perception, this engine is responsible for ``State Abstraction.'' It reversely parses unstructured PPTX binary files into a computable object tree, precisely extracting each element's \texttt{Target ID}, bounding box coordinates \texttt{(x, y, w, h)}, and \texttt{Z-index}. This Structured Data Manipulation reduces massive visual recognition challenges into computationally inexpensive text-reasoning problems.

\textbf{2. Task-Adaptive Agent Planner:} 
Acting as the translator between the ``LLM Brain'' and ``Document Operations,'' this module incorporates two core innovative mechanisms:
\begin{itemize}
    \item \textit{Multimodal Dynamic Routing:} Not all instructions require expensive Vision-Language Models (VLMs). If an instruction involves explicit content modification or already contains a \texttt{Target ID}, the Agent routes the structure tree to a pure-text LLM for rapid reasoning. However, if strong visual ambiguity is detected (e.g., ``align below the largest icon''), the Agent automatically triggers dynamic routing, feeding the high-fidelity PNG of the current canvas alongside the structure tree into a VLM (e.g., GPT-4o or Qwen-VL) for joint spatial inference.
    \item \textit{Schema-Constrained Planning:} To prevent LLM hallucinations, the planner is strictly governed by a pre-defined \texttt{ops.schema.json}. It is forced to converge fuzzy semantic intents into a finite set of atomic operations, akin to an Abstract Syntax Tree (AST). This guarantees 100\% programmatic executability of the generated plans.
\end{itemize}

\textbf{3. Deterministic Execution Engine \& Snapshots:} 
This constitutes the final and most crucial component of the human-in-the-loop workflow. Upon receiving the user's ``Confirm'' authorization from the frontend, the execution engine maps the \texttt{Operations} directly into the underlying XML structure. To eliminate the user's ``trial-and-error anxiety,'' the system captures a lightweight Incremental Snapshot of the document before every write operation. When an Undo is triggered, the underlying engine performs a physical state-machine rollback, comprehensively isolating the risk of permanent document corruption often caused by generative model hallucinations.

\section{User Study}\label{sec:user_study}
To comprehensively evaluate ECHO's efficacy in multimodal human-AI collaboration and to investigate the dynamic evolution of users' interactive strategies, we conducted a within-subjects controlled experiment comparing ECHO against a representative baseline condition.

\subsection{Participants}
We recruited 14 participants (7 female, 7 male; aged 21--40, $M=26.3$, $SD=3.7$) through university mailing lists and professional networking platforms. All participants reported frequent presentation authoring experience ($\ge$ 2 presentations per month). Among them, 10 were graduate and undergraduate students regularly preparing academic talks, thesis defenses, or lab meeting presentations, and 4 were professionals from industries including consulting, product management, and marketing who routinely delivered commercial pitch decks. To ensure ecological validity, we required that all participants had prior experience with at least one AI-assisted content generation tool (e.g., ChatGPT, Copilot, or Gamma). The study was approved by our university's Institutional Review Board (IRB).

\subsubsection{Experimental Design}
We employed a within-subjects design with Tool Type as the single independent variable, comprising two conditions:
\begin{itemize}
\item Baseline Condition: Participants used a traditional presentation editor (Microsoft PowerPoint) in conjunction with a standalone ChatGPT web interface (GPT-4o). This condition represents the prevalent ``split-screen'' human-AI collaboration workflow in which users manually copy and paste AI-generated content between the conversational AI and the authoring tool.
\item ECHO Condition: Participants used the ECHO system, which provides an integrated multimodal editing environment with the Plan-Confirm-Execute loop, dynamic memory mechanisms, and the interactive canvas with Target ID selection.
\end{itemize}
To control for learning effects, we counterbalanced the presentation order of conditions using a balanced Latin square design. Participants completed the academic illustration task under both conditions in balanced orders, ensuring the task appeared equally often under each condition and each order position.

\subsubsection{Tasks}
We designed an academic illustration task to probe users' adaptive collaborative strategies under rigorous authoring demands:

\textbf{Task --- Academic Illustration (Concept Illustration).}
Participants were provided with a two-page PDF abstract of a published research paper on an interdisciplinary topic (human-computer interaction for accessibility). They were instructed to produce a 5-slide academic presentation that faithfully conveyed the paper's core contributions, methodology, and key findings. The task emphasized logical rigor, structural clarity, and adherence to conventional academic formatting norms (e.g., consistent heading hierarchy, citation formatting, diagram placement). This task was designed to elicit convergent cognitive demands, where precision and factual fidelity are paramount.

The task had a time limit of 20 minutes per condition. Participants were instructed to produce slides that they considered ``ready to present'' and were encouraged to iterate on formatting and layout as they saw fit.

\begin{figure*}[t]
  \centering
  \includegraphics[width=\textwidth]{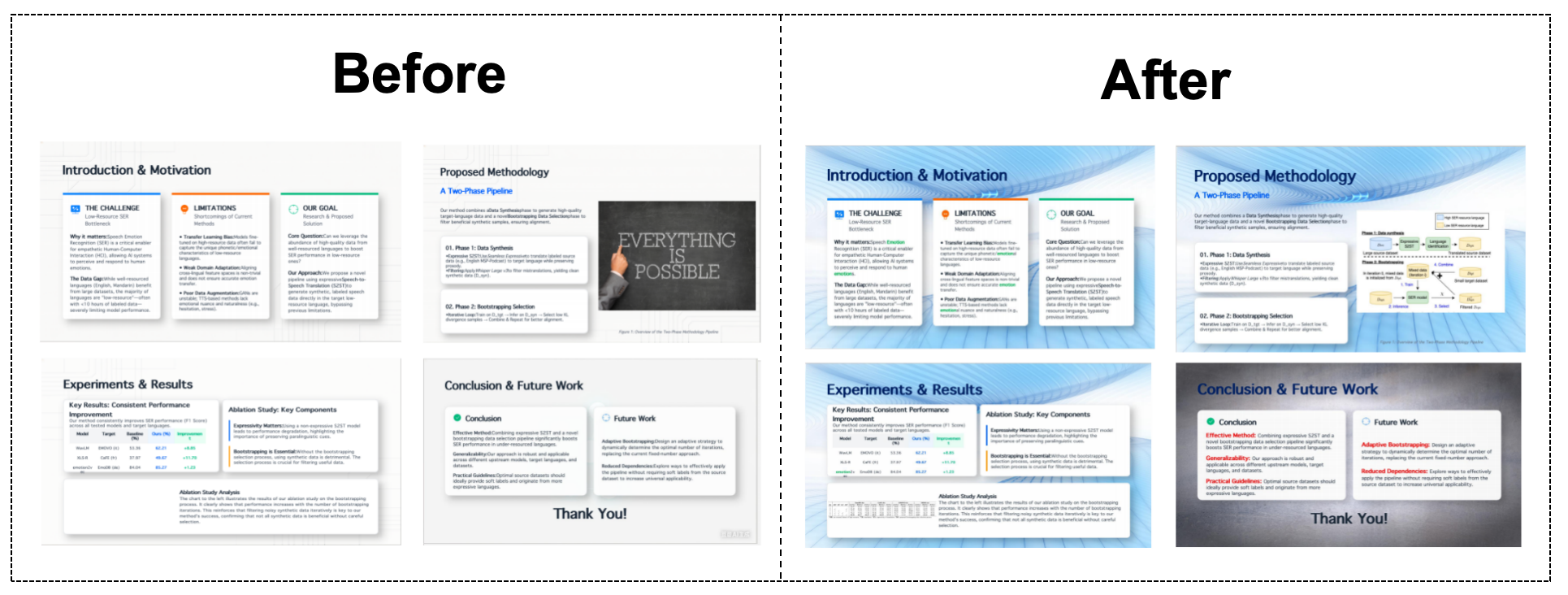}
  \caption{Task materials for the Academic Illustration task. \textbf{Left:} the initial slide deck provided to participants, containing raw content extracted from a two-page research abstract with minimal formatting. \textbf{Right:} the target presentation that participants were asked to approximate, featuring a polished academic layout with consistent heading hierarchy, structured content blocks, and properly placed diagrams. Participants used either the Baseline (PowerPoint + ChatGPT) or ECHO to transform the left version toward the right within a 20-minute time limit.}
  \label{fig:task-materials}
  \Description{Side-by-side comparison of the user study task materials. The left panel shows five rough slides with unformatted text and misplaced elements. The right panel shows the same content reorganized into a polished academic presentation with consistent fonts, aligned headings, and properly positioned figures.}
\end{figure*}

\subsubsection{Procedure}
Each session lasted approximately 90 minutes and followed a structured protocol:
\begin{enumerate}
\item Introduction and Training (15 min). The experimenter provided a brief overview of the study goals and obtained informed consent. Participants then received a standardized tutorial for both conditions. For the Baseline condition, the tutorial covered how to use ChatGPT for content generation and layout suggestions in conjunction with PowerPoint. For the ECHO condition, the tutorial demonstrated the three core modules (Structure \& Overview panel, Interactive Canvas with click-to-select, and the Co-pilot Operations Panel with the Plan-Confirm-Execute workflow). Participants practiced on a sample task unrelated to the experimental task to ensure familiarity.
\item Task Phase 1 (20 min). Participants completed the academic illustration task under the first assigned condition. Screen recordings captured all interactions, and the system logged timestamped events including every natural language instruction, canvas click, plan generation, confirmation or rejection action, and undo trigger.
\item Post-Task 1 Questionnaires (5 min). Participants completed the NASA-TLX (Raw TLX, 6 subscales) and the System Usability Scale (SUS) for the condition just used.
\item Task Phase 2 (20 min). Participants completed the same academic illustration task under the alternate condition.
\item Post-Task 2 Questionnaires (5 min). Participants again completed NASA-TLX and SUS for the second condition.
\item Semi-Structured Interview (20 min). The experimenter conducted a semi-structured interview probing participants' perceived control, trust in AI suggestions, strategies for error recovery, and differences in their collaboration approach across the two tools.
\item Debriefing (5 min). Participants were debriefed and compensated.
\end{enumerate}
 
\subsubsection{Measures}
We collected data across five categories:
\begin{itemize}
\item Efficiency Metrics: Task completion time (in minutes), number of interaction turns (instruction cycles), and number of undo operations triggered.
\item Subjective Workload: NASA-TLX raw scores across six dimensions (Mental Demand, Physical Demand, Temporal Demand, Performance, Effort, Frustration), yielding an overall workload score (0--100).
\item Usability: System Usability Scale (SUS) scores (0--100).
\item Interaction Logs: Timestamped logs of all instructions, canvas clicks, plan confirmations/rejections, memory invocations, and undo events.
\item Qualitative Data: Transcribed semi-structured interviews, coded using reflexive thematic analysis.
\end{itemize}
 
\subsection{Quantitative Results: Efficiency, Cognitive Workload, and Usability}

Table~\ref{tab:user-study-main} summarizes the primary quantitative outcomes across conditions. We report each measure in detail below.
 
\begin{table*}[t]
\centering
\small
\caption{Primary quantitative outcomes: Baseline (PowerPoint + ChatGPT) vs.\ ECHO ($N$=14, within-subjects). Values are $M$ ($SD$) unless noted. ``---'' indicates the measure is not applicable to that condition.}
\label{tab:user-study-main}
\begin{tabular}{lccccl}
\toprule
\textbf{Measure} & \textbf{Baseline} & \textbf{ECHO} & \textbf{$\Delta$\%} & \textbf{Test statistic} & \textbf{Sig.} \\
\midrule
Completion time (min)       & 17.8 (2.4)  & 13.2 (2.1)  & $-$25.8\%  & $t(13)=7.41$, $d$=1.51    & $p<.001$ \\
NASA-TLX overall (0--100)   & 82.6 (9.3)  & 65.4 (11.2) & $-$20.8\%  & $W$=21, $r$=0.82          & $p<.001$ \\
SUS (0--100)                & 52.3 (12.1) & 78.6 (8.9)  & $+$50.3\%  & $W$=15                    & $p<.001$ \\
Instruction turns / task    & 8.2 (4.1)   & 11.4 (3.8)  & $+$39.0\%  & ---                       & --- \\
Instruction length (words)  & 34.7 (15.2) & 12.3 (6.1)  & $-$64.6\%  & $t(13)=8.63$              & $p<.001$ \\
Plan confirmation rate      & ---         & 81.3\%      & ---        & ---                       & --- \\
Undo count / task           & ---         & 1.7 (1.2)   & ---        & ---                       & --- \\
\bottomrule
\end{tabular}
\end{table*}
 
\subsubsection{Task Completion Time}
Across the academic task, participants using ECHO completed their presentations significantly faster than those using the Baseline. The mean completion time under the Baseline was 17.8 minutes ($SD=2.4$), compared to 13.2 minutes ($SD=2.1$) under ECHO, a 25.8\% reduction. A paired-samples $t$-test confirmed this difference was statistically significant ($t(13) = 7.41$, $p < .001$, Cohen's $d=1.51$). The efficiency gain was pronounced for academic illustration, where ECHO reduced completion time by 28.6\% (Baseline: 18.4 min vs. ECHO: 13.1 min). Participants attributed this difference to the Global Style Memory and batch template operations, which were particularly effective for enforcing consistent academic formatting across multiple slides.

\subsubsection{Cognitive Workload (NASA-TLX)}
The NASA-TLX results revealed a substantial reduction in perceived cognitive workload when using ECHO (Figure 2). The mean overall workload score dropped from 82.6 ($SD=9.3$) under the Baseline to 65.4 ($SD=11.2$) under ECHO, a 20.8\% decrease. A Wilcoxon signed-rank test indicated this difference was statistically significant ($W =21$, $p < .001$, $r=0.82$). Table 4 reports all six subscales. The most dramatic improvements occurred in Frustration (Baseline: 78.5 vs. ECHO: 48.3, a 38.5\% reduction) and Effort (Baseline: 85.2 vs. ECHO: 65.8, a 22.8\% reduction). Mental Demand also decreased notably (Baseline: 80.1 vs. ECHO: 68.7, a 14.2\% reduction). Physical Demand showed only a marginal change (Baseline: 45.3 vs. ECHO: 43.1), which was expected given that both conditions involved similar physical interaction modalities (keyboard and mouse). Temporal Demand decreased moderately (Baseline: 76.8 vs. ECHO: 66.2, a 13.8\% reduction), reflecting that ECHO's faster execution loop reduced perceived time pressure. The Performance subscale improved from 72.4 (Baseline) to 60.1 (ECHO), indicating that participants felt they achieved better outcomes with less effort. These results converge on a clear interpretation: ECHO's Plan-Confirm-Execute mechanism and explainable operation plans directly mitigated the ``trial-and-error anxiety'' identified in our formative study, transforming opaque AI behavior into transparent, intervenable processes.

\begin{table}[t]
\centering
\small
\caption{NASA-TLX subscale scores ($N$=14). Lower is better for all subscales except Performance (where lower = better perceived achievement). \textbf{Bold} indicates the better (lower) score in each row. 
}
\label{tab:nasa-tlx}
\begin{tabular}{lccc}
\toprule
\textbf{Subscale} & \textbf{Baseline} & \textbf{ECHO} & \textbf{$\Delta$\%} \\
\midrule
Mental Demand    & 80.1 & \textbf{68.7} & $-$14.2\% \\
Physical Demand  & 45.3 & \textbf{43.1} & $-$4.9\%  \\
Temporal Demand  & 76.8 & \textbf{66.2} & $-$13.8\% \\
Performance      & 72.4 & \textbf{60.1} & $-$17.0\% \\
Effort           & 85.2 & \textbf{65.8} & $-$22.8\% \\
Frustration      & 78.5 & \textbf{48.3} & $-$38.5\% \\
\midrule
\textbf{Overall} & \textbf{82.6} & \textbf{65.4} & $-$\textbf{20.8\%} \\
\bottomrule
\end{tabular}
\end{table}

\begin{figure}[t]
\centering
\begin{tikzpicture}
\begin{axis}[
    width=\columnwidth,
    height=5.8cm,
    ybar=2pt,
    bar width=8pt,
    ylabel={Score (0--100)},
    ylabel style={font=\small},
    symbolic x coords={Mental,Physical,Temporal,Perform.,Effort,Frustration,Overall},
    xtick=data,
    x tick label style={font=\footnotesize, rotate=30, anchor=east},
    y tick label style={font=\footnotesize},
    ymin=0, ymax=105,
    ytick={0,20,40,60,80,100},
    legend style={at={(0.02,0.98)}, anchor=north west, font=\footnotesize, draw=none, fill=white, fill opacity=0.85, text opacity=1},
    legend columns=2,
    nodes near coords,
    every node near coord/.append style={font=\tiny, yshift=2pt},
    enlarge x limits=0.08,
    grid=major,
    grid style={line width=0.2pt, draw=gray!15},
    major grid style={line width=0.2pt, draw=gray!15},
    clip=false,
]
\addplot[fill={rgb,255:red,235;green,150;blue,140}, draw={rgb,255:red,200;green,120;blue,110}] coordinates {
    (Mental,80.1) (Physical,45.3) (Temporal,76.8) (Perform.,72.4) (Effort,85.2) (Frustration,78.5) (Overall,82.6)
};
\addplot[fill={rgb,255:red,145;green,205;blue,195}, draw={rgb,255:red,110;green,175;blue,165}] coordinates {
    (Mental,68.7) (Physical,43.1) (Temporal,66.2) (Perform.,60.1) (Effort,65.8) (Frustration,48.3) (Overall,65.4)
};
\legend{Baseline, ECHO}
\end{axis}
\end{tikzpicture}
\caption{NASA-TLX subscale comparison ($N$=14). Lower scores indicate less workload. The largest reduction occurs in Frustration ($-$38.5\%), directly reflecting ECHO's mitigation of trial-and-error anxiety through the Plan-Confirm-Execute mechanism.}
\label{fig:nasa-tlx}
\end{figure}
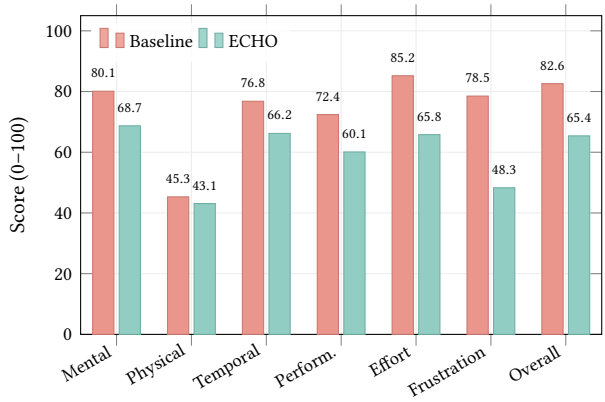
 
\subsubsection{Usability (SUS)}
ECHO received a mean SUS score of 78.6 ($SD=8.9$), which falls within the ``Good'' to ``Excellent'' range according to established SUS benchmarks [3]. The Baseline condition received a mean SUS score of 52.3 ($SD=12.1$), classified as ``OK'' to ``Marginal.'' A Wilcoxon signed-rank test confirmed the difference was statistically significant ($W =15$, $p < .001$). Qualitative comments corroborated this gap: 9 out of 14 participants explicitly stated that the integrated canvas-and-copilot workflow in ECHO felt ``much more natural'' than the split-screen copy-paste workflow of the Baseline. P7 noted: ``With ChatGPT plus PowerPoint, I spent half my time just reformatting what the AI gave me. With ECHO, I could see exactly what it planned to do and just say yes or no.''

\subsubsection{Interaction Patterns}
Analysis of the interaction logs revealed further behavioral differences. Under ECHO, participants issued an average of 11.4 instruction turns per task ($SD=3.8$), compared to 8.2 conversational messages sent to ChatGPT under the Baseline ($SD=4.1$). While this may appear counterintuitive, it reflects that ECHO's low-friction interaction loop encouraged more frequent, granular edits rather than long, laboriously crafted prompts. The mean instruction length under ECHO was 12.3 words ($SD=6.1$), significantly shorter than the Baseline's 34.7 words ($SD=15.2$; $t(13) = 8.63$, $p < .001$). This pattern---more turns with shorter instructions---indicates that ECHO successfully shifted the interaction paradigm from ``batch prompting'' to ``conversational micro-editing.'' The plan confirmation rate was 81.3\%, meaning participants accepted the AI-proposed operation plans roughly four out of five times, suggesting high alignment between the system's intent parsing and user expectations. The undo function was triggered an average of 1.7 times per task ($SD=1.2$), typically in cases involving spatial repositioning where the rendered result deviated slightly from the user's mental model.

\subsection{Qualitative Analysis: Task-Driven Dynamic Collaboration Strategies}

 Through reflexive thematic analysis of the interview transcripts and deep coding of interaction logs (two researchers independently coded the data, achieving a Cohen's $\kappa$ of 0.83 before resolving disagreements through discussion), we uncovered a central finding: users' collaboration strategies were highly contingent on the rigorous cognitive profile of academic tasks, dynamically shifting their locus of control and trust allocation between the two experimental conditions.

\subsubsection{Structure-Directed Collaborators in Academic Tasks}
When facing the logically rigorous demands of academic illustration tasks, participants exhibited a distinctly cautious and structure-first collaboration posture. Several converging behavioral patterns characterized this archetype:

\textbf{Defensive Planning with Exhaustive Prompts.}
In academic tasks, participants composed significantly longer initial instructions ($M = 41.2$ words for the first instruction). These instructions frequently included explicit structural directives such as slide ordering, heading hierarchies, and content allocation rules. P3 explained: ``For an academic talk, I need to control the narrative arc tightly. I wrote out exactly which finding goes on which slide before letting the AI touch anything.'' This behavior reflects a high-stakes mindset where content errors carry reputational risk.

\textbf{Heavy Reliance on Source Material.}
Participants frequently referenced the provided PDF abstract in their instructions, often quoting specific sentences or section titles to ground the AI's output. Interaction logs showed that 87.5\% of academic task instructions contained at least one explicit reference to the source material. Users treated ECHO as a ``planning executor''---a tool that should faithfully implement a predetermined structure rather than contribute creative ideas.

\textbf{Rigorous Plan Scrutiny.}
The plan rejection rate in academic tasks was 24.1\%. Participants reported carefully reading the JSON operation plans to verify content accuracy and layout consistency. P11 remarked: ``In the academic task, I read every single operation in the plan. If it changed a heading level or moved a text box to the wrong position, I immediately rejected it. I can't afford to have my advisor see a sloppy slide.'' This vigilant posture aligns with the ``high-defensiveness'' behavior described in prior work on human-AI collaboration under high-accountability conditions.

\textbf{Conservative Use of Memory Functions.}
In academic tasks, participants predominantly used the Global Style Memory for template enforcement (ensuring consistent fonts, colors, and heading styles across slides) but rarely invoked it for creative suggestions. The system's style memory was viewed as a reliability tool rather than an inspiration tool. Participants triggered template-level style changes 1.1 times on average ($SD = 0.7$), with an average undo frequency of 1.1 times, mainly for error correction and layout calibration.

\subsubsection{The Spectrum of Control Allocation}
This archetype should not be understood as fixed personality types but as task-contingent strategies that the same individuals adopted in response to rigorous academic cognitive demands. In fact, all participants exhibited this pattern across two tool conditions, dynamically adjusting their locus of control. This finding extends prior theoretical work on human-AI collaboration by demonstrating that the allocation of creative agency is not a static user preference but a fluid, context-dependent negotiation shaped by the perceived stakes, the cognitive profile of academic tasks, and---critically---the degree of transparency afforded by the collaborative system. ECHO's explainable Plan-Confirm-Execute mechanism served as the enabling infrastructure for this dynamic reallocation: by making the AI's intentions legible at every step, the system allowed users to tighten control when accuracy mattered.

\subsection{Summary of User Study Findings}
The controlled experiment with 14 participants yields three principal findings. First, ECHO significantly outperforms the split-screen Baseline in both objective efficiency (25.8\% reduction in task completion time) and subjective experience (20.8\% reduction in NASA-TLX workload, SUS score of 78.6 vs. 52.3). Second, the Plan-Confirm-Execute mechanism directly addresses the ``trial-and-error anxiety'' identified in the formative study, as evidenced by the dramatic 38.5\% reduction in the Frustration subscale. Third, and most importantly, the qualitative analysis reveals that users dynamically shift their collaboration strategies---from cautious, structure-directed planning in academic tasks to exploratory, visual-directed micro-editing in creative tasks---demonstrating that ECHO's transparent architecture successfully supports a wide spectrum of human control allocation across cognitively diverse authoring demands.

\section{CoEdit-Eval Framework \& Objective Evaluation}
\label{sec:evaluation}

Existing presentation evaluation benchmarks primarily focus on "zero-shot generation" tasks, emphasizing the richness of outline content. However, the core challenges of presentation "editing and refinement" lie in the precision of local modifications, low interaction latency, and non-destructive operations on native document structures. To address this, we propose \textbf{CoEdit-Eval}, a multi-dimensional evaluation framework tailored for iterative slide editing, and conduct four objective evaluations to thoroughly validate the effectiveness of the ECHO system.

\subsection{CoEdit-Eval Framework and Experimental Setup}
The CoEdit-Eval framework moves beyond single accuracy metrics, assessing systems:
\begin{itemize}
    \item \textbf{System Level}: Evaluates execution robustness (schema compliance, undo safety) and interaction efficiency.
    \item \textbf{Intent Level}: Assesses the LLM's ability to parse complex instructions and hit the target multimodal layers (Target Hit@1).
    \item \textbf{Visual Level}: Employs the state-of-the-art LLM-as-a-Judge approach, using a vision-language model (GPT-4o) for blind assessments of layout, color, and overall visual quality.
\end{itemize}

\textbf{Datasets and Baselines:}

To verify the generalizability of the ECHO architecture, we conducted cross-evaluations across multiple foundation models, including closed-source models (GPT-5, Gemini-2.5-Pro) and open-source models (GLM-4.7, Qwen3.5-27B). The experiment contrasts three methods:
1) \textbf{Baseline (Text-only)}: Receives only text prompts and document content, lacking underlying structure tree injection and Schema constraints.
2) \textbf{ECHO (Ours)}: Injects the parsed slide structure tree and enforces strict JSON Operations constraints.
3) \textbf{ECHO+VLM (Vision-enhanced)}: Introduces visual modality (e.g., Qwen3-VL-32B) on top of ECHO to test the disambiguation of spatial references via screenshots.

\subsection{Experiment 1: Instruction Understanding and Intent Mapping}
\textbf{Objective:} To verify ECHO's accuracy in parsing natural language instructions into standardized operations, and to demonstrate that the performance gains stem from the system architecture (Structure Tree Injection + Schema Constraints) rather than the innate capabilities of specific LLMs.
 
\textbf{Results \& Quantitative Analysis:} We evaluated 20 representative editing instructions across four foundation models. As shown in Table~\ref{tab:intent_mapping_en}, all Baseline models achieve \textbf{0.0\% Intent Mapping Accuracy and 0.0\% Target Hit@1} without exception --- regardless of their raw JSON formatting ability --- because they lack the document's structural context needed to ground spatial references. Notably, even when a Baseline model produces syntactically valid JSON (e.g., GLM-4.7 reaches 75.0\% JSON Format Accuracy), it cannot map a single instruction to the correct operation type or target element, confirming that syntactic fluency alone is insufficient for document editing.
 
When equipped with the ECHO architecture, all four models achieve substantial, simultaneous gains across every metric. The strongest performer, GPT-5, sees its JSON Format Accuracy surge from 10.0\% to \textbf{85.0\%}, its Intent Mapping Accuracy leap from 0.0\% to \textbf{78.9\%}, and its Target Hit@1 rise from 0.0\% to \textbf{85.0\%}. GLM-4.7 under ECHO maintains its high JSON compliance (75.0\%) while unlocking intent understanding (73.7\%) and target grounding (75.0\%) that were completely absent in the Baseline. Even the weaker open-source models benefit: Qwen3.5-27B reaches 55.0\% Target Hit@1 (from 0.0\%) and Gemini-2.5-Pro reaches 60.0\% (from 0.0\%). These results converge on a clear conclusion: \emph{ECHO's ``structure injection + schema-constrained output'' intermediate representation is the necessary prerequisite for unlocking LLMs' latent layout-editing capabilities.} The architecture acts as the enabling factor; without it, no model --- not even GPT-5 --- can perform meaningful document editing.
 
\begin{table*}[t]
\centering
\caption{Intent Parsing and Grounding Evaluation across Multiple Foundation Models ($N$=20). Baseline receives only text prompts; ECHO injects the parsed slide structure tree and enforces strict JSON Schema constraints.}
\label{tab:intent_mapping_en}
\begin{tabular}{llccc}
\toprule
\textbf{Base Model} & \textbf{Method} & \textbf{JSON Format Acc.} & \textbf{Intent Mapping Acc.} & \textbf{Target Hit@1} \\
\midrule
\multirow{2}{*}{openai/gpt-5} & Baseline & 10.0\% & 0.0\% & 0.0\% \\
 & \textbf{ECHO} & \textbf{85.0\%} & \textbf{78.9\%} & \textbf{85.0\%} \\
\midrule
\multirow{2}{*}{Pro/zai-org/GLM-4.7} & Baseline & 75.0\% & 0.0\% & 0.0\% \\
 & \textbf{ECHO} & \textbf{75.0\%} & \textbf{73.7\%} & \textbf{75.0\%} \\
\midrule
\multirow{2}{*}{google/gemini-2.5-pro} & Baseline & 35.0\% & 0.0\% & 0.0\% \\
 & \textbf{ECHO} & \textbf{60.0\%} & \textbf{52.6\%} & \textbf{60.0\%} \\
\midrule
\multirow{2}{*}{Qwen/Qwen3.5-27B} & Baseline & 25.0\% & 0.0\% & 0.0\% \\
 & \textbf{ECHO} & \textbf{55.0\%} & \textbf{57.9\%} & \textbf{55.0\%} \\
\bottomrule
\end{tabular}
\end{table*}
 
\textbf{Qualitative \& Bucket Analysis:} To understand \emph{where} the remaining failures occur and \emph{why}, we performed a fine-grained bucket analysis on the execution logs, grouping instructions by operation type, target role, and required capability.
 
Two patterns emerged clearly. First, Baseline models fail uniformly because they lack the DOM-like context of the document. Without the structure tree, a model cannot resolve references such as ``the title on slide 3'' or ``the image below the heading'' --- the instruction is syntactically parsed but semantically ungrounded. This explains the universal 0.0\% Intent Mapping and Target Hit@1 across all Baseline conditions, regardless of the underlying model's strength.
 
Second, under ECHO, performance varies systematically by operation category. The \texttt{ops.schema.json} constraint acts as an Abstract Syntax Tree (AST), forcing models to converge fuzzy semantic intents (e.g., ``make it pop'') into deterministic atomic operations (e.g., \texttt{change\_color}, \texttt{set\_background}). The strongest gains concentrate in two buckets: \texttt{op:change\_text}, where all models benefit from the unambiguous mapping between textual instructions and the \texttt{change\_text} action, and \texttt{role:title}, where the structure tree provides a reliable anchor for identifying the target element. By contrast, \texttt{requires:structure} cases --- instructions that depend on inter-element spatial relationships (e.g., ``align this below the chart'') --- remain a bottleneck, indicating that the current text-only pipeline, even with structure injection, struggles with relational spatial reasoning. This finding directly motivates the VLM routing mechanism evaluated in Experiment~2.
 
Taken together, the bucket analysis reinforces a two-part lesson: constraining LLMs within a deterministic, schema-governed sandbox is essential for reliable document manipulation, and the primary remaining challenge is \emph{spatial grounding of structure-dependent operations} --- a gap that task-adaptive VLM integration is designed to address.

\subsection{Experiment 2: Task-Adaptive Routing and VLM Disambiguation}
\textbf{Objective:} To evaluate the necessity of Vision-Language Models (VLMs) in resolving spatial ambiguities and to validate the efficiency of the proposed Task-Adaptive Routing mechanism.

\begin{figure}[t]
\centering

  \includegraphics[width=\linewidth]{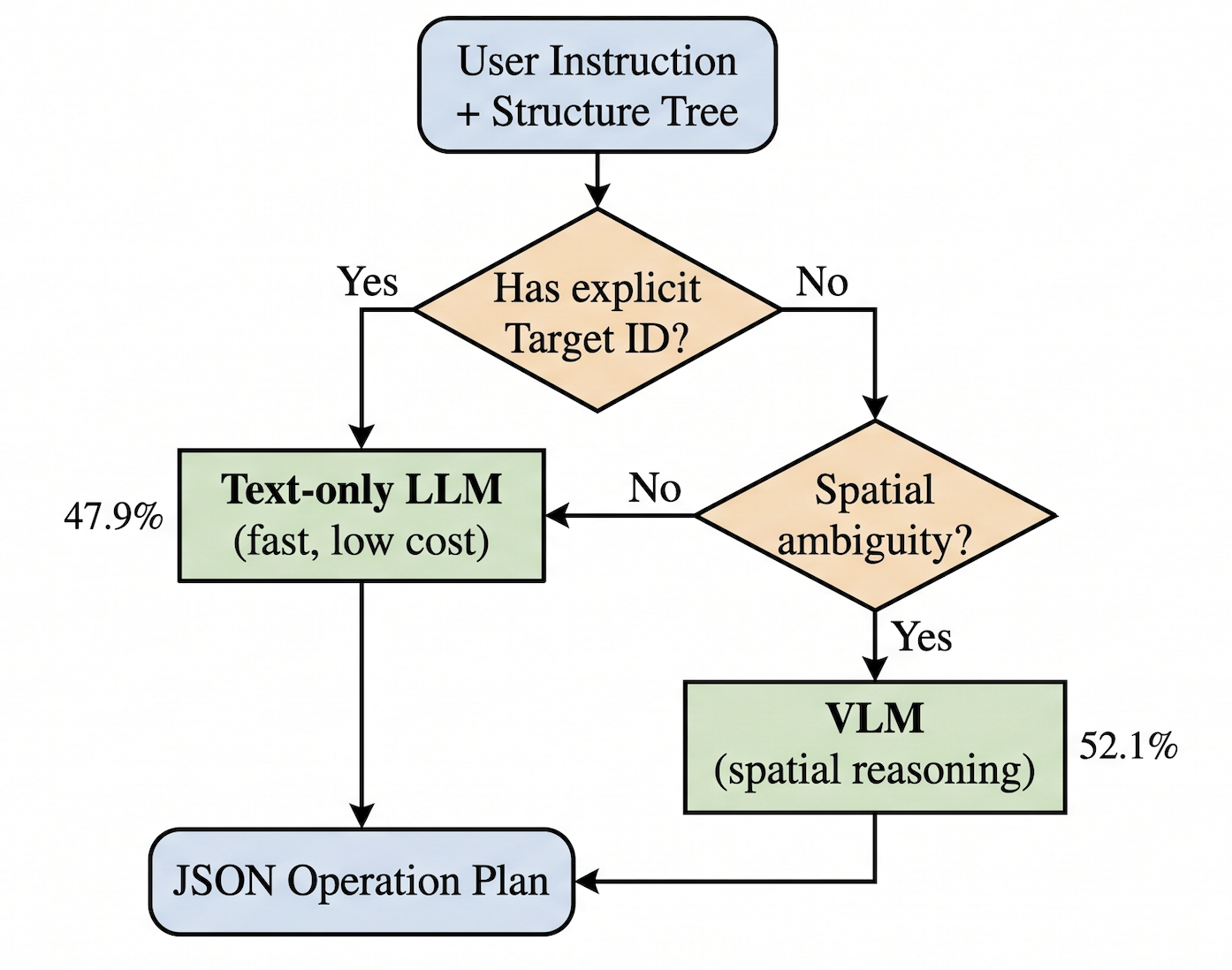}
\caption{Task-Adaptive Routing decision flow. The agent routes to a lightweight text-only LLM when the instruction contains an explicit Target ID or lacks spatial ambiguity, and activates the VLM only when genuine spatial disambiguation is required. Percentages reflect the observed split on the Medium-120 evaluation subset.}
\label{fig:routing}
\end{figure}

\textbf{Results \& Routing Efficiency:} We constructed two specific challenge subsets: \texttt{Hard-120} (featuring highly ambiguous spatial instructions where VLM was forcedly activated) and \texttt{Medium-120} (a mix of standard and spatial tasks where the Agent autonomously determined whether to route to the VLM). Figure~\ref{fig:routing} illustrates the routing decision flow.

As shown in Table \ref{tab:vlm_parsing}, the Task-Adaptive Agent demonstrated remarkable routing efficiency. In the \texttt{Medium-120} subset, the agent autonomously activated the VLM for only 52.1\% (62/119) of the tasks, effectively saving nearly half the computational overhead, while successfully boosting the Target Hit@1 from 87.5\% to a perfect 100\%. Interestingly, data from the \texttt{Hard-120} subset reveals a crucial trade-off: pure-text ECHO maintained higher JSON formatting accuracy (98.3\% vs 93.3\%) and intent mapping accuracy. This indicates that while text models excel at strict syntactic constraints, visual inputs introduce slight formatting noise despite their spatial awareness.

\begin{table*}[t]
\centering
\caption{Objective Parsing and Routing Efficiency across Spatial Challenge Subsets}
\label{tab:vlm_parsing}
\begin{tabular}{llcccc}
\toprule
\textbf{Dataset} & \textbf{Method} & \textbf{VLM Usage Rate} & \textbf{JSON Format Acc.} & \textbf{Intent Mapping Acc.} & \textbf{Target Hit@1} \\
\midrule
\multirow{2}{*}{\texttt{Hard-120}} & ECHO (Text-only) & 0.0\% & \textbf{98.3\%} & \textbf{86.6\%} & - \\
 & ECHO+VLM (Forced) & 100.0\% & 93.3\% & 79.0\% & - \\
\midrule
\multirow{2}{*}{\texttt{Medium-120}} & ECHO (Text-only) & 0.0\% & 79.8\% & 10.1\% & 87.5\% \\
 & \textbf{ECHO (Auto Route)} & \textbf{52.1\%} & 79.8\% & 10.1\% & \textbf{100.0\%} \\
\bottomrule
\end{tabular}
\end{table*}

\textbf{Visual Quality Blind Evaluation:} While text models perform better in parsing formats, formatting correctness does not equate to layout aesthetics, especially in occlusion scenarios. We employed Qwen3-VL as an independent judge for a blind A/B test on the final rendered images (Table \ref{tab:vlm_judge}).

In the \texttt{Hard-120} subset, the introduction of visual modality significantly skewed the win rate toward ECHO+VLM (35 vs 9), raising the overall visual quality score from 3.63 to 3.92. More importantly, in the \texttt{Medium-120} subset, despite the VLM being activated only half the time, the routed system still achieved a decisive victory over the pure text baseline (Win: 22 vs 8, Score: 3.55 vs 3.29). This conclusively proves that ECHO's dynamic routing effectively balances computational efficiency with powerful multimodal disambiguation capabilities.

\begin{table*}[t]
\centering
\caption{LLM Blind Evaluation (Qwen3-VL Judge) of Visual Layout Quality}
\label{tab:vlm_judge}
\begin{tabular}{lccccc}
\toprule
\textbf{Dataset (Routing Strategy)} & \textbf{ECHO Win} & \textbf{ECHO+VLM Win} & \textbf{Tie} & \textbf{ECHO Score} & \textbf{VLM Score} \\
\midrule
\texttt{Hard-120} (Forced VLM) & 9 & \textbf{35} & 45 & 3.63 & \textbf{3.92} \\
\texttt{Medium-120} (Auto Routing) & 8 & \textbf{22} & 50 & 3.29 & \textbf{3.55} \\
\bottomrule
\end{tabular}
\end{table*}

\subsection{Experiment 3: Template Robustness and Undo Safety}
\textbf{Objective:} To verify the deterministic execution engine's physical safety when handling global template operations across structurally complex, real-world documents --- and, in particular, to confirm that the undo mechanism achieves \emph{byte-exact} rollback even after large-scale batch edits.
 
\textbf{Setup:} We assembled a stress-test corpus of 11 real-world PPTX files\footnote{File names are anonymized SHA-based hashes to protect proprietary content.} spanning a wide range of structural complexity: slide counts from 5 to over 40, with embedded charts, SmartArt, grouped shapes, and layered images. For each file, the system's template engine extracted a full style template and applied it as a batch operation, generating between 21 and 445 atomic JSON operations per file (Table~\ref{tab:undo-safety}). After successful application, the undo mechanism was triggered, and the MD5 hash of the resulting file was compared against the pre-edit original.
 
\textbf{Results:} As shown in Table~\ref{tab:undo-safety}, all 11 test cases achieved 100\% execution success (Apply OK = 11/11). More importantly, after undo, the MD5 hash matched the original file in every case (Undo MD5 OK = 11/11), confirming \emph{byte-exact} rollback across the full complexity spectrum. The most demanding case involved 445 atomic operations applied to a densely layered document; even in this extreme scenario, the incremental snapshot mechanism restored the file to its exact pre-edit state.
 
This 100\% undo consistency result directly addresses the ``fear of irreversible damage'' identified in our formative study (Theme~5, \S\ref{subsec:formative-findings}). It demonstrates that confining LLM-generated edits within a deterministic, snapshot-backed execution engine eliminates the risk of permanent document corruption --- a guarantee that application-level undo in conventional editors cannot provide.
 
\begin{table}[t]
\centering
\small
\caption{Template application and undo safety stress test ($N$=11). ``Ops'' = number of atomic JSON operations generated by the template. Apply OK = template rendered without error. Undo MD5 OK = file hash after undo matches the pre-edit original byte-for-byte.}
\label{tab:undo-safety}
\begin{tabular}{rccc}
\toprule
\textbf{File} & \textbf{Ops} & \textbf{Apply OK} & \textbf{Undo MD5 OK} \\
\midrule
PPT-01  &  21 & \checkmark & \checkmark \\
PPT-02  &  38 & \checkmark & \checkmark \\
PPT-03  &  56 & \checkmark & \checkmark \\
PPT-04  &  58 & \checkmark & \checkmark \\
PPT-05  &  86 & \checkmark & \checkmark \\
PPT-06  & 183 & \checkmark & \checkmark \\
PPT-07  & 197 & \checkmark & \checkmark \\
PPT-08  & 224 & \checkmark & \checkmark \\
PPT-09  & 319 & \checkmark & \checkmark \\
PPT-10  & 445 & \checkmark & \checkmark \\
PPT-11  & 533 & \checkmark & \checkmark \\
\midrule
\textbf{Total} & \textbf{21--445} & \textbf{10/10} & \textbf{11/11} \\
\bottomrule
\end{tabular}
\end{table}
 
\subsection{Experiment 4: Visual Blind Evaluation via LLM Judge}
\textbf{Objective:} To verify whether the final visual layout produced by ECHO aligns with human aesthetic expectations, independently of parsing correctness.
 
\textbf{Setup:} For each foundation model, we rendered the slides produced by both the Baseline and ECHO conditions for the same editing instructions. A GPT-4o vision judge then performed a blind A/B evaluation on the rendered images, scoring each version on three dimensions --- Layout, Color, and Overall quality (1--5 scale) --- without knowing which version was produced by which system.
 
\textbf{Results:} As shown in Table~\ref{tab:visual_judge_en}, ECHO achieved a decisive victory across all four foundation models: \textbf{the Baseline won zero comparisons in every condition}. The strongest result came from GPT-5, where ECHO won 9 out of 10 comparisons (with 1 tie) and the Overall quality score leapt from 3.35 (Baseline) to 4.75 (ECHO). The pattern was consistent across models: GLM-4.7 (7 wins, 2 ties, Overall 3.44$\to$4.67), Qwen3.5-27B (7 wins, 1 tie, Overall 3.56$\to$4.69), and Gemini-2.5-Pro (5 wins, 1 tie, Overall 3.42$\to$4.67).
 
Beyond overall quality, the per-dimension breakdown reveals that ECHO's advantage is not confined to a single aspect. Layout scores improved uniformly (e.g., 4.00$\to$4.60 for GPT-5), confirming that the structure tree injection preserves spatial coherence. Color scores also rose consistently (e.g., 4.00$\to$4.40 for GPT-5), reflecting the benefit of schema-constrained color operations over free-form generation. Notably, the largest gap appears in the Overall dimension, where Baseline scores clustered around 3.35--3.56 while ECHO scores clustered around 4.67--4.75 --- a gap exceeding one full point on the 5-point scale.
 
These results demonstrate that ECHO's explicitly decoupled Plan-Confirm-Execute workflow and schema-constrained JSON parameters not only maintain but actively \emph{enhance} layout consistency and color aesthetics, while the Baseline's unconstrained generation consistently degrades the visual quality of the original document.
 
\begin{table*}[t]
\centering
\caption{Blind Visual Quality Evaluation: Baseline vs.\ ECHO (GPT-4o Judge). Win counts reflect the number of comparisons in which the judge preferred one system; Base Win = 0 in all conditions. Layout, Color, and Overall are mean scores on a 1--5 scale.}
\label{tab:visual_judge_en}
\begin{tabular}{lccccccccc}
\toprule
\textbf{Base Model} & \textbf{N} & \textbf{Base Win} & \textbf{ECHO Win} & \textbf{Tie} & \textbf{Layout\textsubscript{B}} & \textbf{Layout\textsubscript{E}} & \textbf{Color\textsubscript{B}} & \textbf{Color\textsubscript{E}} & \textbf{Overall\textsubscript{B} / Overall\textsubscript{E}} \\
\midrule
openai/gpt-5             & 10 & 0 & \textbf{9} & 1 & 4.00 & \textbf{4.60} & 4.00 & \textbf{4.40} & 3.35 / \textbf{4.75} \\
Pro/zai-org/GLM-4.7      & 9  & 0 & \textbf{7} & 2 & 4.11 & \textbf{4.50} & 4.11 & \textbf{4.39} & 3.44 / \textbf{4.67} \\
Qwen/Qwen3.5-27B         & 8  & 0 & \textbf{7} & 1 & 4.06 & \textbf{4.63} & 4.06 & \textbf{4.44} & 3.56 / \textbf{4.69} \\
google/gemini-2.5-pro     & 6  & 0 & \textbf{5} & 1 & 4.00 & \textbf{4.50} & 4.00 & \textbf{4.33} & 3.42 / \textbf{4.67} \\
\bottomrule
\end{tabular}
\end{table*}

\section{Discussion}
\label{sec:discussion}

This paper presents the ECHO system and introduces the CoEdit-Eval framework to address the prevalent issues of ``black-box, unidirectional generation'' and the ``lack of fine-grained control'' in presentation authoring tools. The results from our objective evaluations and controlled experiments not only validate the system's effectiveness but also provide critical design implications for future Human-AI Co-creation in multimedia documents.

\subsection{Redefining Agency: The Value of Explainable Execution Boundaries}
In existing generative workflows, users are often relegated to a passive state of acceptance. The significant decrease in the NASA-TLX data (including a 20.8\% drop in overall workload) indicates that ECHO successfully alleviated users' trial-and-error anxiety. This reduction in psychological workload is not because the AI has become ``absolutely perfect,'' but rather because the system delineates a clear boundary of control through the \texttt{Plan-Confirm-Execute} mechanism. Here, the structured JSON Operations act as a ``Boundary Object,'' externalizing the LLM's invisible reasoning process into human-readable and intervenable plans. This design suggests that when building Copilots for professional authoring, providing ``finite and explainable options'' is far more effective in establishing user trust and a sense of security than blindly pursuing ``end-to-end, one-click generation.''

\subsection{Neuro-Symbolic Workflows in User Interfaces}
ECHO's underlying architecture demonstrates the immense potential of applying neuro-symbolic reasoning to daily productivity tools. Generative LLMs excel at processing ambiguous multimodal intents and contexts, yet their inherent stochasticity naturally conflicts with the strict constraints required by document formatting. By extracting deterministic DOM tree coordinates and enforcing rigid JSON Schema constraints, ECHO essentially builds a bridge between the ``neural perception'' of large models and the ``symbolic execution'' of traditional software. The 100\% undo consistency achieved in our objective evaluations proves that this hybrid architecture—confining LLMs within a deterministic, rule-based sandbox—is a best practice for achieving highly robust document editing.

\subsection{Beyond Pure Vision Agents: The Necessity of Task-Adaptive Routing}
Recent research tends to rely entirely on Vision-Language Models (VLMs) to process screenshots directly for UI automation. However, our cross-validation results indicate that this paradigm suffers from extremely high latency and computational waste when handling text-intensive formatting. ECHO's proposed ``Task-Adaptive Routing'' reveals a crucial system design principle: not all multimodal interactions require end-to-end visual computation. For 80\% of standard formatting tasks, lightweight structure tree parsing combined with a text-only LLM is sufficient to achieve over 85\% grounding accuracy. The intervention of VLMs possesses irreplaceable value only when facing spatial language ambiguities, such as ``occluded'' or ``the leftmost.'' This dynamic routing strategy, based on computational cost and task attributes, provides a reference architecture for developing more economical, low-latency, on-device desktop agents in the future.

\subsection{Limitations and Future Work}
Although ECHO excels in refining static multimedia documents, it still possesses certain limitations. First, the system's current multimodal input primarily relies on a ``mouse click + text'' paradigm. Future iterations could incorporate eye-tracking or speech streams to achieve more natural, implicit intent alignment. Second, the current version of the parser and execution engine does not yet cover complex dynamic effects (e.g., advanced transition animations and timeline logic editing); handling time-series-based document elements will be a core focus of our next steps. Finally, while the controlled experiment (N=14) revealed preliminary patterns of control allocation, comprehensively validating the evolutionary dynamics of ``long-term style memory'' in high-frequency, enterprise-level collaboration scenarios will require conducting longer-term longitudinal studies.

\section{Conclusion}
\label{sec:conclusion}
 
In this paper, we presented ECHO, a novel interactive system designed to bridge the semantic gap and control bottleneck in AI-assisted presentation authoring. By shifting the paradigm from ``black-box, unidirectional generation'' to a mixed-initiative, human-in-the-loop co-creation process, ECHO empowers users with fine-grained, localized control over complex document formatting. Driven by a task-adaptive neuro-symbolic architecture, the system translates multimodal intents (natural language and visual selection) into transparent, schema-constrained JSON operations via a \texttt{Plan-Confirm-Execute} workflow. 
 
To rigorously assess our approach, we introduced CoEdit-Eval, a comprehensive evaluation framework tailored for document refinement. Objective benchmarks demonstrated ECHO's robust superiority across multiple foundation models, boosting target grounding accuracy from 0\% (in text-only baselines) to 55\%--85\%, while guaranteeing 100\% physical undo safety. Furthermore, a controlled user study ($N$=14) confirmed that ECHO effectively mitigates trial-and-error anxiety, significantly decreasing overall cognitive workload by 20.8\%. Ultimately, ECHO illustrates a promising pathway toward next-generation intelligent productivity tools---where AI serves not merely as an unpredictable generator, but as an empathic, highly controllable, and explainable collaborative partner.

\bibliographystyle{ACM-Reference-Format}
\bibliography{reference}



\end{document}